\documentclass[11pt]{article}

\usepackage[a4paper,margin=1in]{geometry}
\usepackage{amsmath,amssymb}
\usepackage{graphicx}
\usepackage{booktabs}
\usepackage{array}
\usepackage{float}
\usepackage{caption}
\usepackage{authblk}
\usepackage{microtype}
\usepackage[dvipsnames]{xcolor}
\usepackage{titlesec}
\usepackage[hidelinks,colorlinks=true,linkcolor=RoyalBlue,citecolor=RoyalBlue,urlcolor=RoyalBlue,bookmarks=false,pdfpagemode=UseNone]{hyperref}

\definecolor{accent}{HTML}{2F5FE0}
\titleformat{\section}{\large\bfseries\color{accent!85!black}}{\thesection}{0.6em}{}
\titleformat{\subsection}{\normalsize\bfseries}{\thesubsection}{0.6em}{}
\newcommand{\T}{^{\mathsf{T}}}
\newcommand{\E}{\mathbb{E}}

\title{\textbf{Innovating Risk Modelling for Global Funds.}\\[4pt]
{\large\normalfont A Principal-Component Engine With an AI Interpretation Layer\\[1pt]
for a Benchmark-less Global Innovation Fund}}

\author{Swaraj Gambhir}
\author{Thanu George}
\author{Kairavi Sivasankar}
\affil{\small GenInnov Pte. Ltd.}
\date{\today}

\begin{document}
\maketitle

\begin{abstract}
\noindent
Markowitz defined portfolio risk as an \emph{internal} property, built from the covariance among a
book's own holdings rather than the distance to any index. Seventy years of simplification reversed
that. The market beta of CAPM, the fixed style and industry axes of Barra-type models, and the
promotion of benchmark deviation to the definition of risk all traded the inward view for an
\emph{external} one. Risk became \emph{distance from an index}. For a fund that fits no benchmark, that
trade fails. A global book concentrated in a few markets and a few innovation sectors has no natural
index to deviate from, and the active-risk number it produces measures the mismatch, not the risk. We
return to the covariance. Principal component analysis (PCA) recovers the systematic structure inside
the portfolio directly from its own returns. PCA has always carried one cost: its factors resist a
plain-English reading. We clear that with a generative-AI labelling layer. It names the leading
factors, ranked by their actual contribution to risk rather than by universe variance, and a
deterministic rubric keeps it from inventing structure the loadings do not contain. Around this sit
four independent signals. Density-based clustering with a \emph{mismatch ratio} flags groups whose
risk outruns their capital. A sign-invariant PCA Risk Score (PRS) marks the names that build the
dominant factor bets. A standalone \emph{Bleed} score catches the slow capital destroyers PCA cannot
see. A trailing-return timing gate routes disagreements between the risk signals and recent price
action to human judgment. We run the full engine on a proxy global-innovation book of thirty names
over one year.
\end{abstract}

\noindent\fbox{\parbox{\linewidth}{\small\textbf{Data disclaimer.} All holdings, weights, and
figures in this paper are computed on an \emph{illustrative proxy} global-innovation portfolio
constructed for demonstration. They are not GenInnov's live book and are not investment advice.
The contribution of this paper is the risk-measurement \emph{engine}, not the example portfolio it
runs on.}}

\section{Introduction: the seventy-year drift}
For a portfolio with weights $w=(w_1,\dots,w_N)$ over $N$ assets with covariance matrix $\Sigma$,
Markowitz~\cite{markowitz1952} defined risk as
\begin{equation}
\sigma_p^2 = w\T \Sigma\, w = \sum_i \sum_j w_i w_j \sigma_{ij},
\label{eq:markowitz}
\end{equation}
where the off-diagonal terms $\sigma_{ij}$ ($i\neq j$) are where diversification lives. Crucially,
there is no index in \eqref{eq:markowitz}: risk is an \emph{inward-looking} property of the holdings
themselves. The model was, however, computationally heavy (the number of distinct covariance terms
grows as $N(N-1)/2$), and every major development since can be read as a way of avoiding the full
matrix.

CAPM~\cite{sharpe1964,lintner1965,mossin1966} assumed a single market portfolio, collapsing the
matrix to $\sigma_{ij}=\beta_i\beta_j\sigma_m^2$ and risk to one beta per asset. Barra's multi-factor
models~\cite{barra1998} replaced the one axis with a fixed set of externally
specified style and industry axes, $\Sigma = X F X\T + D$. Finally, the active-management framework
of Grinold and Kahn~\cite{grinoldkahn} applied the Markowitz variance machinery to the deviation
portfolio $w_p-w_b$ and promoted the result, \emph{tracking error}, to the headline definition of risk,
\begin{equation}
\mathrm{TE} = \sqrt{(w_p-w_b)\T \Sigma\, (w_p-w_b)}, \qquad
\mathrm{IR} = \frac{\E[R_p - R_b]}{\mathrm{TE}}.
\label{eq:te}
\end{equation}
The cumulative result is an industry in which risk is, by default, distance from an index, and in
which the further a portfolio sits from a benchmark, the ``riskier'' it is declared to be, regardless
of how sound its internal diversification is.

\paragraph{Why this fails for a benchmark-less fund.} A truly global innovation fund fits no benchmark category. The
standard axes (global vs.\ regional, emerging vs.\ developed, value vs.\ growth) do not describe a
book that is global but concentrated in a few markets and thematic across a handful of
innovation-driven sectors. When $w_b$ is an ill-fitting index, $w_p-w_b$ in \eqref{eq:te} is
dominated by the trivial fact that the fund holds what the index does not; the resulting active-risk
number measures the mismatch, not the genuine risk inside the book. Table~\ref{tab:bench} makes this
concrete: forced against five plausible indices, the same proxy book reports tracking errors from
$12.6\%$ to $30.1\%$ and information ratios that swing from $+1.32$ to $-2.60$ purely with the choice
of yardstick. None of these numbers describes the portfolio; they describe the gap to an index it
was never built to resemble.

\begin{table}[t]
\centering\small
\caption{Benchmark-relative statistics for the proxy book against five indices. The wide,
sign-flipping spread is the symptom: benchmark-relative risk reports the choice of index, not the
portfolio. (Window: 2025-06-01 to 2026-05-31.)}
\label{tab:bench}
\begin{tabular}{lrrrrr}
\toprule
Benchmark & $\beta$ & Alpha (ann.) & Tracking err. & Info.\ ratio (ann.) & Down-capture \\
\midrule
Nasdaq Composite & 0.75 & $+12.9\%$ & $13.9\%$ & $+0.37$ & $72.1\%$ \\
S\&P 500         & 0.97 & $+15.5\%$ & $13.9\%$ & $+1.06$ & $91.4\%$ \\
KOSPI (Korea)    & 0.26 & $+6.5\%$  & $30.1\%$ & $-2.60$ & $23.1\%$ \\
TAIEX (Taiwan)   & 0.50 & $+0.3\%$  & $18.0\%$ & $-2.03$ & $53.8\%$ \\
MSCI World       & 1.19 & $+12.9\%$ & $12.6\%$ & $+1.32$ & $112.3\%$ \\
\bottomrule
\end{tabular}
\end{table}

\paragraph{Beta, alpha, and the information ratio against indices the fund does not resemble.} The numbers can
still be computed, of course. Such a book can be run against MSCI World, the MSCI All Country World Index,
the Nasdaq-100, the S\&P~500, an emerging-markets index, or a thematic-growth basket and read off a
$\beta$, an alpha, a tracking error and an information ratio for each; the machinery never refuses. But
the figures are unstable and uninformative. A portfolio can co-move with an index it barely overlaps,
because both load on the same broad macro forces, so each regression captures a different blend of
shared exposure: the same book shows $\beta = 0.97$ against one plausible index and $\beta = 0.26$
against another (Table~\ref{tab:bench}). No single reading can stand as the portfolio's market sensitivity. Alpha behaves the same way. It is
only the intercept of an index-dependent regression, and it moves with the yardstick, not the book.
Tracking error restates the distance to the chosen index, and the information ratio rescales that
distance. Change the index and every figure moves. The book's risk has not changed; the reference
point has. What comes out is a tidy dashboard of precise numbers that says which index the book resembles least. It does not, however, make any inference about where
the real risk is in the book, nor does it answer the question that matters: which position, if
trimmed, would actually lower the risk being carried?

The industry did not go back to first principles, but instead corrected each piece of the link between
CAPM and Grinold--Kahn by adding another outside axis of measurement. This means the factor zoo of
Fama, French, and Carhart~\cite{carhart1997,famafrench2015}, as well as a performance-ratio
tradition~\cite{sortino1994,sharpe1994} that captured risk in after-the-fact scores. All these are
external yardsticks.

The present research, however, returns to the fundamentals and numerically measures the systematic
structure of the portfolio. This work is motivated by PCA, a method that detects the underlying
structure that gives rise to the co-movement rather than imposing it~\cite{pearson1901,hotelling1933}.
Statistical factor models have been in the toolkit for decades, from Ross's arbitrage pricing
theory~\cite{ross1976} to Avellaneda and Lee's PCA eigenportfolios~\cite{avellaneda2010}. But they have
never been the language a manager speaks to a client. The
eigenvectors of PCA are hard to explain, having earned a plain-English reading only on the yield
curve~\cite{littermanscheinkman}. That communication gap, not a mathematical flaw, is the obstacle a generative-AI
layer now removes.

\section{The principal-component engine}
\label{sec:engine}
\subsection{From a messy book to a clean return matrix}
Preprocessing is four moves: reconstruct the book as USD weights on a common basis; strip the cash
sleeve (a spurious zero-variance asset); place every holding on one shared business-day calendar
across the US, Korea, Taiwan, Hong Kong, Japan, and Europe (interpolating gaps of at most three
business days; longer gaps exclude the observation); and convert
prices to simple returns $R_{i,t}=P_{i,t}/P_{i,t-1}-1$. The URTH ETF, which tracks
the MSCI World index, is retained as the market proxy for the residualisation.

\subsection{Residual, cross-sectionally-demeaned, covariance PCA}
\label{sec:respca}
Following the residual-PCA methodology of Avellaneda and Lee~\cite{avellaneda2010}, we do not run PCA on raw returns. First we residualise each stock against the market, keeping only
the part that is not the global ``everything moves together'' factor:
\begin{equation}
\varepsilon_{i,t} = R_{i,t} - \big(\alpha_i + \beta_i R_{\mathrm{URTH},t}\big),
\qquad \beta_i = \frac{\mathrm{Cov}(R_i,R_{\mathrm{URTH}})}{\mathrm{Var}(R_{\mathrm{URTH}})}.
\end{equation}
Without this step PC1 would just be market beta, saying nothing about how the holdings differ.
Second, we use cross-sectional demeaning at each date to eliminate any remaining common drift,
\begin{equation}
\tilde{\varepsilon}_{i,t} = \varepsilon_{i,t} - \tfrac{1}{N}\sum_{j}\varepsilon_{j,t}.
\end{equation}
The order matters: residualise \emph{then} demean. Finally we eigendecompose the sample covariance of
the cleaned matrix $X=[\tilde{\varepsilon}_{i,t}]$,
\begin{equation}
\Sigma = \tfrac{1}{T-1} X\T X = V\Lambda V\T = \sum_k \lambda_k v_k v_k\T,
\qquad \mathrm{evr}_k = \frac{\lambda_k}{\sum_j \lambda_j},
\end{equation}
where $X$ is the $T\times N$ matrix of cleaned residuals (dates in rows, assets in columns).
We use the covariance (not the correlation) so each name keeps its actual residual volatility when
portfolio weights are later projected onto the factors.

\subsection{Risk contribution versus explained variance}
\begin{figure}[t]
\centering
\includegraphics[width=0.95\linewidth]{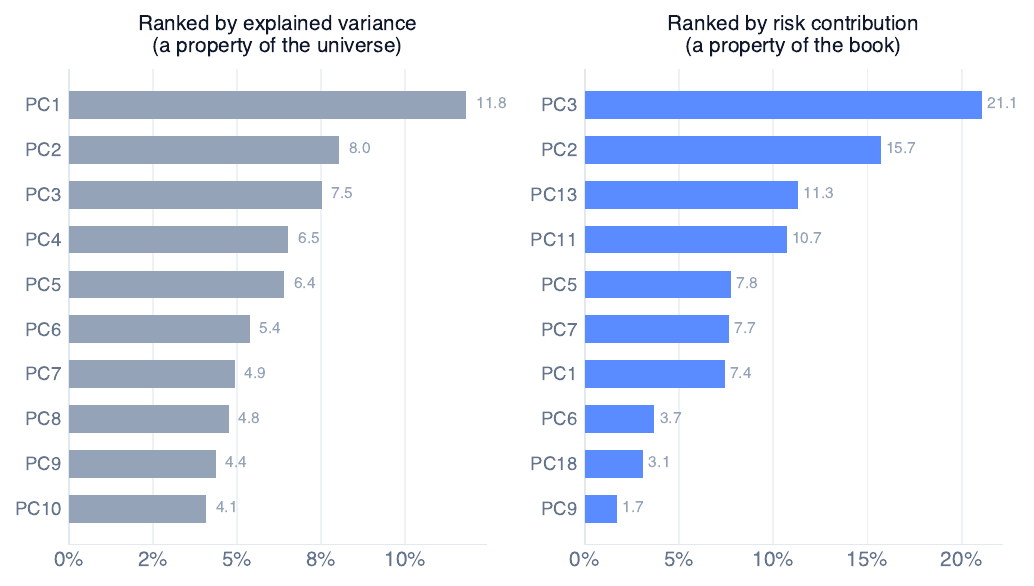}
\caption{The universe's biggest factor is not the book's biggest risk. Left: principal components
ranked by explained-variance ratio. Right: the same components ranked by portfolio risk
contribution \eqref{eq:rc}. The order changes once the actual weights are applied.}
\label{fig:evr}
\end{figure}

Explained variance is a property of the universe. A manager cares how much each factor drives the
variance in \emph{this} book. The key identity is obtained by substituting the
eigendecomposition into $\sigma_p^2 = w\T\Sigma\, w$, an exact, cross-term-free decomposition:
\begin{equation}
\sigma_p^2 = \sum_k \lambda_k (v_k\T w)^2,
\qquad
\mathrm{RC}_k = \lambda_k (v_k\T w)^2,
\qquad
\%\mathrm{RC}_k = \frac{\mathrm{RC}_k}{\sigma_p^2},
\qquad
\sum_k \%\mathrm{RC}_k = 1.
\label{eq:rc}
\end{equation}
Throughout \S\ref{sec:engine}, $\Sigma$ is the cleaned residual covariance of \S\ref{sec:respca},
so \eqref{eq:rc} decomposes the book's \emph{residual} variance, not the total variance a client sees;
\S\ref{sec:clusters} deliberately reverts to the raw covariance when budgeting realised risk.
The projection $v_k\T w$ is the book's net exposure to factor $k$. Because it enters squared, the
ranking by risk contribution often differs substantially from the ranking by explained
variance: a factor that is fourth or fifth by universe variance can be first by contribution to the
book. Figure~\ref{fig:evr} shows this re-ordering on the proxy book: PC3 is third by
explained variance but first by risk contribution, while PC13 and PC11 leap up the ranking once the
weights are applied. Explained variance answers ``what moves the market of these stocks?''; risk
contribution answers ``what moves this book?'' We rank by the latter everywhere.

\subsection{Three loadings, and the sign caveat}
\begin{figure}[ht]
\centering
\begin{minipage}[c]{0.5\linewidth}
  \centering
  \includegraphics[width=\linewidth]{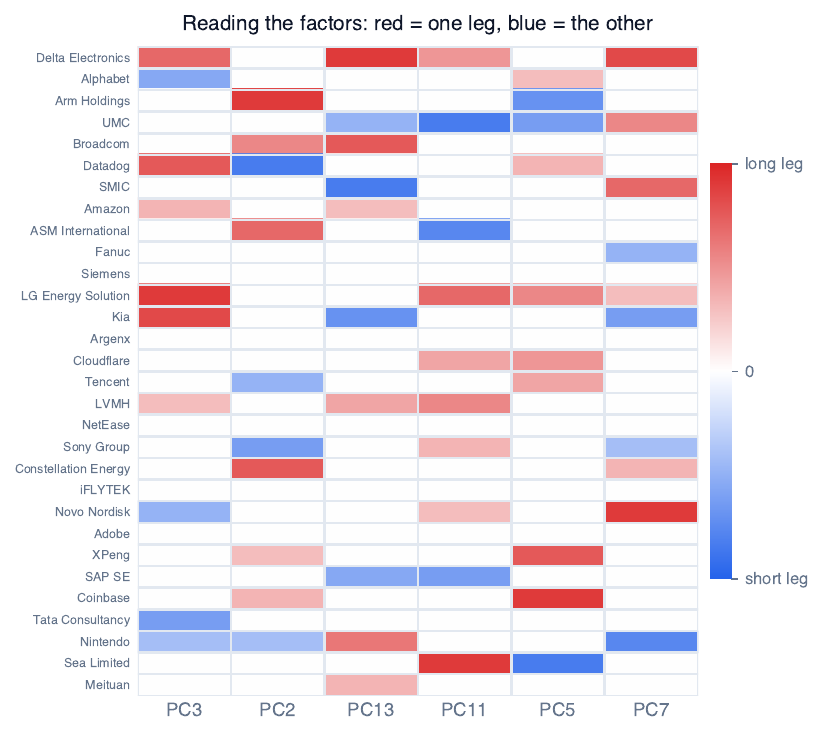}
\end{minipage}\hfill
\begin{minipage}[c]{0.47\linewidth}
  \centering\scriptsize
  \setlength{\tabcolsep}{3pt}\renewcommand{\arraystretch}{1.15}
  \begin{tabular}{@{}l>{\raggedright\arraybackslash}p{0.58\linewidth}r@{}}
    \toprule
    PC & Factor & \%RC \\
    \midrule
    PC3  & EV Supply Chain Momentum vs.\ Diversified Global Defensives            & 21.1 \\
    PC2  & Semiconductor Infrastructure vs.\ Consumer Software \& Entertainment   & 15.7 \\
    PC13 & Taiwan--US Hardware Alpha vs.\ Greater China Foundry \& Mobility        & 11.3 \\
    PC11 & SEA E-commerce Growth vs.\ Semi-Cap Equipment Cyclicals                & 10.7 \\
    PC5  & Crypto/Speculative Asset vs.\ Broad Tech/E-commerce                    & 7.8 \\
    PC7  & Healthcare Innovation vs.\ Japan Consumer/Industrial Cyclicals         & 7.7 \\
    \bottomrule
  \end{tabular}
\end{minipage}
\caption{Reading the factors. \emph{Left:} for the top six risk-contributing components, the holdings
that define each leg, shaded by a correlation-style loading proxy (red = one leg, blue = the other);
columns are keyed by PC index. \emph{Right:} the AI-generated name for each factor, in descending order
of contribution to portfolio risk (\%RC). Factor names are produced by the labelling layer of
\S\ref{sec:ai}.}
\label{fig:load}
\end{figure}

The same decomposition yields three loading forms: the raw eigenvector $v_{i,k}$ (a unit-norm
direction, the only form in which the variance algebra closes exactly), the scaled loading
$v_{i,k}\sqrt{\lambda_k}$ (covariance units), and the correlation-style loading
\begin{equation}
\ell_{i,k}^{\mathrm{corr}} = \frac{v_{i,k}\sqrt{\lambda_k}}{\sigma_i} \in [-1,1],
\end{equation}
where $\sigma_i$ is the residual volatility of stock $i$ computed from the cleaned matrix $X$ of
\S\ref{sec:respca}; with that definition the $[-1,1]$ bound holds up to numerical error. This form
approximates the correlation of stock $i$ with factor $k$ and is what we hand to humans and to
the labelling model (Figure~\ref{fig:load}). An eigenvector is defined only up to sign: $v_k$ and
$-v_k$ are both valid, and across runs the sign can flip with no economic meaning. Only the
\emph{grouping} of names is invariant, a hard constraint that dictates the sign-invariant design of the
PRS in \S\ref{sec:prs}.

\subsection{AI labels: clearing the century-old interpretation problem}
\label{sec:ai}
A factor called ``PC7'' means nothing to a manager. We pass the top ten PCs \emph{by risk
contribution} (together with their loadings, dominant sectors, and the portfolio's exposure to
each) through an automated labelling pipeline that returns, for every factor, a structured label, a
rationale, a confidence score, and supporting evidence, keyed to a timestamped run identifier. The
pipeline is constrained by a deterministic rubric so that names follow the geometry of the loadings
rather than free invention: a factor that dominates the book's risk (large $\%\mathrm{RC}$) is cast
as a two-sided regime, \emph{``[Style A] vs.\ [Style B]''}, naming the long and short legs read off
the loading signs; a factor carrying one dominant loading well above the rest is cast as
\emph{``Idiosyncratic: [Ticker]''}; the remainder are named for their dominant sector and leading
contributors. The numbers stay fully computed; only the \emph{names} are generated. On the proxy book
the leading factor by risk contribution ($21.1\%$ of portfolio variance, PC3) is labelled
\emph{``EV Supply Chain Momentum vs.\ Diversified Global Defensives''} (long LG Energy
Solution/Kia/Delta Electronics against Tata Consultancy/Alphabet/Novo Nordisk), turning a raw
eigenvector into a sentence a portfolio manager can act on (Figure~\ref{fig:load}).

\section{Clustering and the cluster-level risk budget}
\label{sec:clusters}
PCA tells us \emph{what} drives risk but not \emph{which group of names} to look at first. We cluster
the cleaned residual matrix with HDBSCAN~\cite{mcinnes2017} (density-based, \texttt{min\_cluster\_size=3}) after
denoising the correlation matrix and detoning it, that is, removing the dominant market
mode~\cite{laloux1999,plerou1999}, in the manner standardised by López de
Prado~\cite{lopezdeprado2018,lopezdeprado}. On
the proxy book this returns five dense themes (\emph{AI Compute \& Power}; \emph{Asia EV, Power \&
Robotics}; \emph{Growth Software \& Crypto}; \emph{China Internet \& Tech}; and \emph{Global Mega-Cap
Leaders}) plus a handful of idiosyncratic ``lone wolves'' (the HDBSCAN noise bucket).

\begin{figure}[t]
\centering
\includegraphics[width=0.9\linewidth]{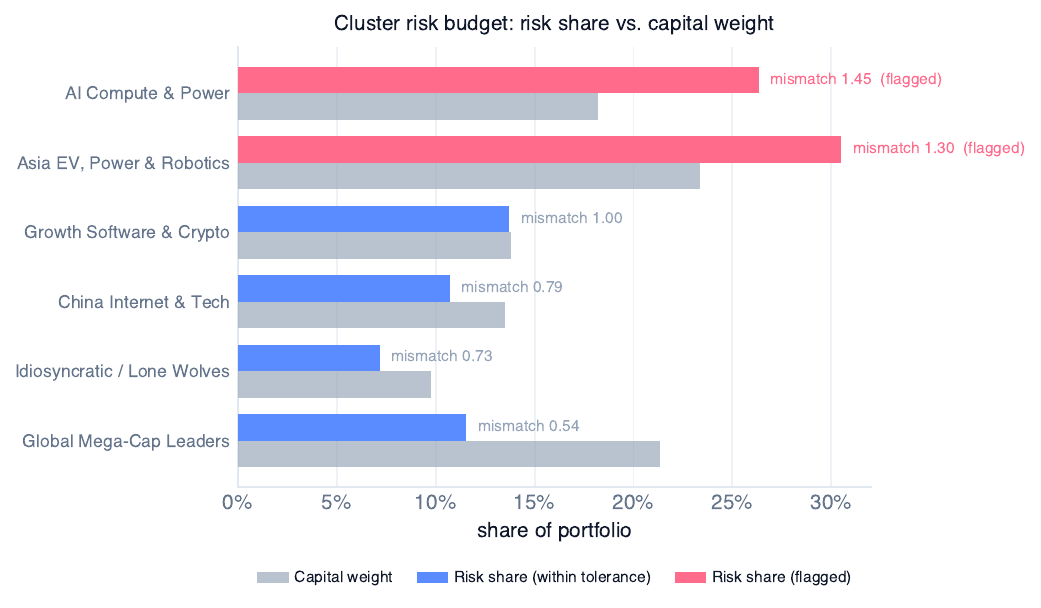}
\caption{Cluster risk budget. Each theme's risk share versus its capital weight, with the mismatch
ratio annotated. Red bars are flagged as over-risked: they carry risk out of proportion to their capital.}
\label{fig:clusters}
\end{figure}

For each cluster we compute the marginal and component contributions to risk and normalise by capital.
Because this budget describes realised risk the client bears, here $\Sigma$ is the raw
(total-return) covariance rather than the residual covariance of \S\ref{sec:respca}:
\begin{equation}
\begin{gathered}
\mathrm{MCR}_i = \frac{(\Sigma w)_i}{\sigma_p}, \qquad
\mathrm{CCR}_i = \frac{w_i (\Sigma w)_i}{\sigma_p}, \qquad
\%\mathrm{RC}_i = \frac{\mathrm{CCR}_i}{\sigma_p} = \frac{w_i (\Sigma w)_i}{\sigma_p^2}, \\[4pt]
\text{mismatch} = \frac{\text{risk share}}{\text{capital weight}}.
\end{gathered}
\end{equation}
A mismatch above $1$ implies that the cluster has more risk exposure relative to its allocation, and
hence the cluster is over-risked and we flag it. The mismatch is the right signal, not the raw risk
share. A large cluster carrying proportionally large risk is balanced; a small cluster carrying
outsized risk is a concentrated bet. Figure~\ref{fig:clusters} presents the budget of the proxy book, where two clusters,
\emph{AI Compute \& Power} and \emph{Asia EV, Power \& Robotics}, have a higher level of risk exposure
than capitalization (mismatch $1.45$ and $1.30$).

\section{Two per-stock signals: PRS and Bleed}
\subsection{The PCA Risk Score (PRS)}
\label{sec:prs}
Within a flagged cluster we need a per-stock number that says \emph{how much does this name build the
portfolio's dominant factor bets?} The PCA Risk Score restricts the risk-contribution machinery to
the top-3 factors by $\mathrm{RC}_k$:
\begin{equation}
\mathrm{PRS}_i = \frac{1}{\sigma_p^2} \sum_{k \in \text{top-3 RC}} \lambda_k\, (v_k\T w)\, w_i\, v_{i,k}.
\label{eq:prs}
\end{equation}
The signed $\mathrm{PRS}_i$ is then percentile-ranked across holdings to a $0$--$100$ column. Two
properties earn \eqref{eq:prs} its form. \emph{It is sign-invariant}: the eigenvector appears twice
(once in $v_k\T w$, once in $v_{i,k}$), so under $v_k\to -v_k$ the product is unchanged, verified
numerically to machine precision. \emph{It is a principled slice of CCR}: summing the summand over
\emph{all} $k$ recovers the textbook component risk share exactly, with respect to the residual
covariance of \S\ref{sec:respca}; the cluster budget of \S\ref{sec:clusters} uses the raw covariance,
and the two lenses are deliberately distinct. Restricting to the top-3
risk-contributing factors is therefore not an approximation but precisely the ``contribution through your
dominant factors'' subset that a concentration-trim decision asks for.

\subsection{The Bleed score: what PCA cannot see}
A stock in a long fundamental decline is invisible to PCA and PRS, or worse, flattering. Variance is
symmetric, so a smooth grind-down reads as low risk; an idiosyncratic decliner lands in its own factor
or the noise bucket, where it reads as \emph{diversifying}. Statistical independence from the factors is not fundamental quality. Hence we include a
stand-alone measure of the stock using just its price series: the Martin (Ulcer Performance)
ratio~\cite{martin1989}:
\begin{equation}
\mathrm{Ulcer}_i = \sqrt{\tfrac{1}{T}\sum_t \mathrm{drawdown}_{t}^2}, \qquad
\mathrm{AnnRet}_i = \Big(\tfrac{P_{\mathrm{end}}}{P_{\mathrm{start}}}\Big)^{252/T}\!\!-1, \qquad
\mathrm{Martin}_i = \frac{\mathrm{AnnRet}_i}{\mathrm{Ulcer}_i}.
\end{equation}
The denominator of the Ulcer incorporates both the depth and duration of underwater periods; a low or
negative Martin ratio is the hallmark of the serial bleeder. We flip and rank it to a $0$--$100$
\emph{Bleed} score that runs on the same scale and in the same direction as the PRS. Independence from
the PRS is by design, since it asks a different question.

\subsection{The timing gate}
Ultimately we have raw $1$-month and $6$-month returns sitting next to the scores, specifically left
out of any composite. An aggressively rising trim candidate or a declining add candidate is sent to
human discretion; the timing gate highlights the contradictions between the risk indicator and the price
action.

\section{The decision table}
\begin{figure}[t]
\centering
\includegraphics[width=0.86\linewidth]{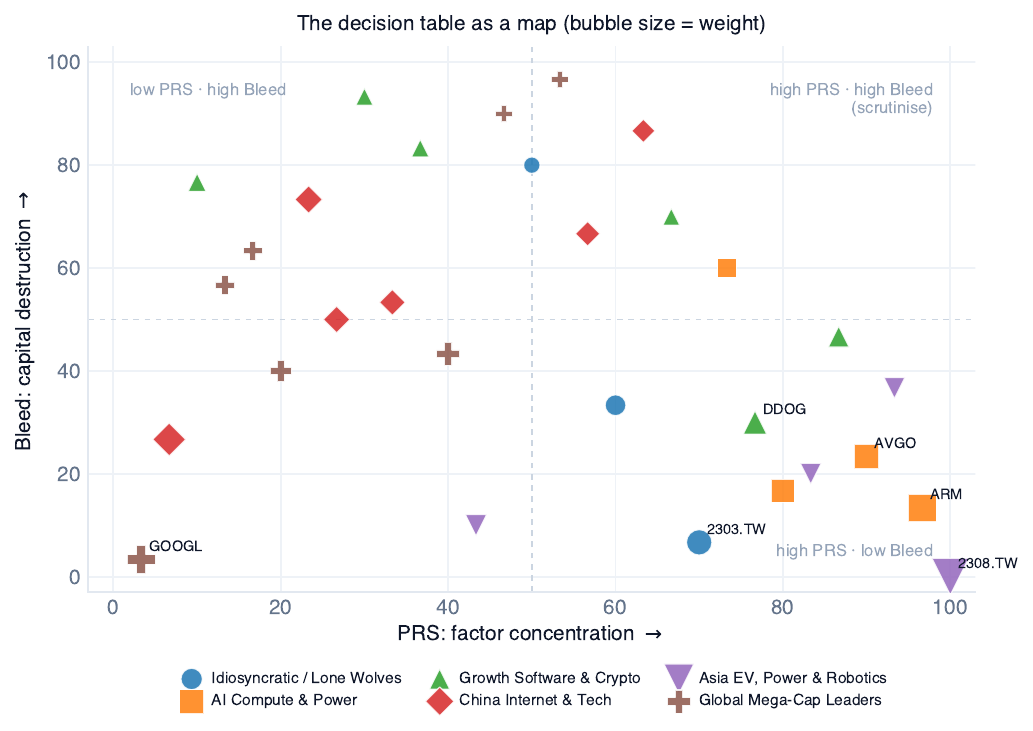}
\caption{The decision table as a map. Each holding by PRS (factor-risk concentration) versus Bleed
(own-price capital destruction), coloured by cluster, sized by weight. The two scores are built from
disjoint information (portfolio covariance structure versus own-price history), so the four quadrants
carry distinct, actionable meanings.}
\label{fig:decision}
\end{figure}

The signals are combined into a single artifact, a decision table with one row per stock containing
its cluster, PRS, Bleed, and trailing returns, read under a cluster $\to$ mismatch $\to$ drill-down
workflow. This is shown in Figure~\ref{fig:decision} as a map, with the concentrated factor builders
(trim candidates that are not themselves bleeding) in the high-PRS / low-Bleed quadrant, and the slow
capital destroyers that PCA alone would miss in the low-PRS / high-Bleed quadrant. Largest positions
are labelled; bubble size is weight in capital.

The two levels are chained together in one workflow. The first gate is the cluster mismatch ratio: a
balanced or risk-diluting cluster is left alone, while a flagged, over-risked cluster is opened up
name by name. Within it, the quadrant map sorts candidates (a high-PRS, low-Bleed name concentrates
the dominant factor bets and is a trim candidate, whereas a low-PRS, low-Bleed name with conviction
is an add candidate), and the raw trailing-return timing gate routes any signal/price disagreement to
human judgment rather than a mechanical trade. The engine narrows the whole book to the few names
that genuinely warrant a decision.

\section{Result on the proxy book}
\begin{figure}[H]
\centering
\includegraphics[width=0.92\linewidth]{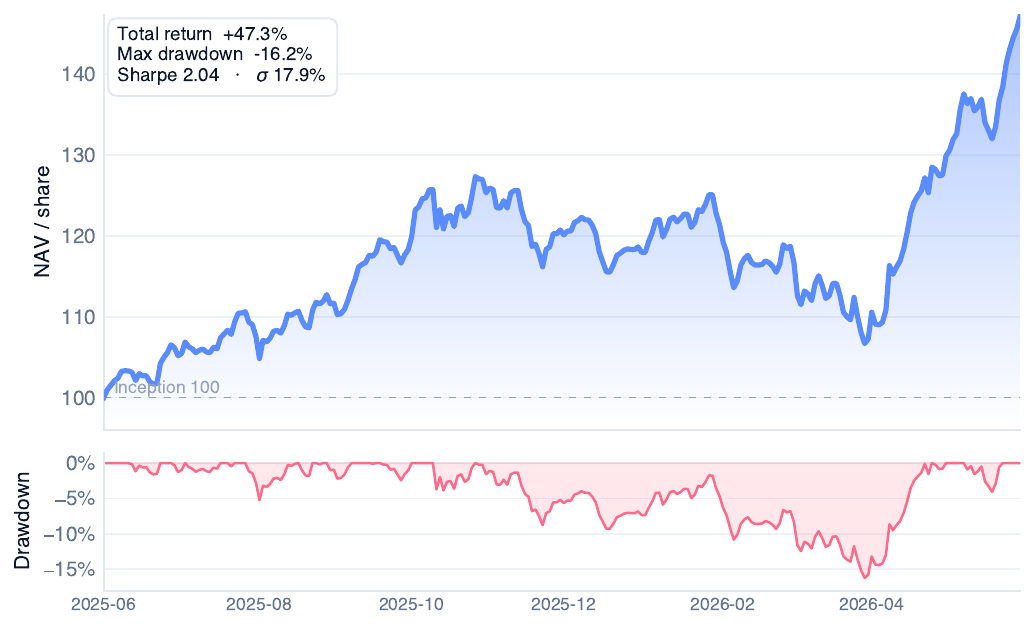}
\caption{Proxy-book NAV per share over the one-year window (top) and drawdown path (bottom). Summary
statistics are computed from the book's own return series, with no benchmark imposed.}
\label{fig:nav}
\end{figure}

Over the one-year window the proxy book returns $+47.3\%$, with a maximum drawdown of $-16.2\%$, an
annualised volatility of $17.9\%$, a Sharpe ratio of $2.04$, and a Sortino ratio of $2.06$
(Figure~\ref{fig:nav}). The Sortino ratio uses a risk-free target, $\tau=r_f=2.77\%$ annualised; at
$\approx1$ bp per day that places it within $\sim$1\% of a zero-threshold semi-deviation. Daily $95\%$
VaR and CVaR are $1.78\%$ and $2.26\%$: losses of at least $1.78\%$ on the worst $5\%$ of days,
averaging $2.26\%$ within that tail. None of these numbers needed a benchmark. In risk terms, the thirty names resolve to a small set of
factor bets. The top three risk-contributing components account for roughly half of portfolio variance.
Two clusters carry most of the variance and most of the downside on a minority of the capital. That is
a statement about the portfolio's own structure. It is the inward-looking measurement
\eqref{eq:markowitz} that benchmark-relative practice set aside, now stated plainly.

\section{Limitations}
The engine measures risk. It does not build or improve the portfolio. PCA on a single window depends
on the sample, and eigenvector stability worsens when eigenvalues are close. This is one reason every
downstream signal is designed to be sign-invariant or standalone. The AI labels follow a fixed guide
and are ranked by computed risk contribution; however, they are still just a way to communicate
numbers, not a source of them. The proxy book is just an example. On a live book, the same pipeline
runs without changes, but the specific clusters, factors, and scores will vary. Finally, PCA is one lens among several. In production, it works alongside traditional exposure measures,
cross-correlation structure, antifragility and convexity checks, tail and stress analysis, and a
monthly risk review of the entire book. Two extensions are planned: moving PRS from a drill-down score
to a full selection axis along with cluster mismatch and Bleed, and adding a conviction-scoring layer
that weighs risk flags based on fundamental thesis strength. PCA is where the measurement starts, not
where it ends.



\begin{thebibliography}{99}
\small
\bibitem[1]{markowitz1952} H.~Markowitz. Portfolio Selection. \textit{Journal of Finance}, 7(1):77--91, 1952.
\bibitem[2]{sharpe1964} W.~F. Sharpe. Capital Asset Prices: A Theory of Market Equilibrium Under Conditions of Risk. \textit{Journal of Finance}, 19(3):425--442, 1964.
\bibitem[3]{lintner1965} J.~Lintner. The Valuation of Risk Assets and the Selection of Risky Investments in Stock Portfolios and Capital Budgets. \textit{Review of Economics and Statistics}, 47(1):13--37, 1965.
\bibitem[4]{mossin1966} J.~Mossin. Equilibrium in a Capital Asset Market. \textit{Econometrica}, 34(4):768--783, 1966.
\bibitem[5]{barra1998} BARRA, Inc. \textit{United States Equity Model Version 3 (USE3): Risk Model Handbook}. BARRA, Inc., 1998.
\bibitem[6]{ross1976} S.~A. Ross. The Arbitrage Theory of Capital Asset Pricing. \textit{Journal of Economic Theory}, 13(3):341--360, 1976.
\bibitem[7]{littermanscheinkman} R.~Litterman and J.~Scheinkman. Common Factors Affecting Bond Returns. \textit{Journal of Fixed Income}, 1(1):54--61, 1991.
\bibitem[8]{sortino1994} F.~A. Sortino and L.~N. Price. Performance Measurement in a Downside Risk Framework. \textit{Journal of Investing}, 3(3):59--64, 1994.
\bibitem[9]{sharpe1994} W.~F. Sharpe. The Sharpe Ratio. \textit{Journal of Portfolio Management}, 21(1):49--58, 1994.
\bibitem[10]{carhart1997} M.~M. Carhart. On Persistence in Mutual Fund Performance. \textit{Journal of Finance}, 52(1):57--82, 1997.
\bibitem[11]{grinoldkahn} R.~C. Grinold and R.~N. Kahn. \textit{Active Portfolio Management}, 2nd~ed. McGraw-Hill, 2000.
\bibitem[12]{avellaneda2010} M.~Avellaneda and J.-H. Lee. Statistical Arbitrage in the US Equities Market. \textit{Quantitative Finance}, 10(7):761--782, 2010.
\bibitem[13]{famafrench2015} E.~F. Fama and K.~R. French. A Five-Factor Asset Pricing Model. \textit{Journal of Financial Economics}, 116(1):1--22, 2015.
\bibitem[14]{mcinnes2017} L.~McInnes, J.~Healy, and S.~Astels. hdbscan: Hierarchical Density Based Clustering. \textit{Journal of Open Source Software}, 2(11):205, 2017.
\bibitem[15]{pearson1901} K.~Pearson. On Lines and Planes of Closest Fit to Systems of Points in Space. \textit{Philosophical Magazine}, 2(11):559--572, 1901.
\bibitem[16]{hotelling1933} H.~Hotelling. Analysis of a Complex of Statistical Variables Into Principal Components. \textit{Journal of Educational Psychology}, 24:417--441, 1933.
\bibitem[17]{laloux1999} L.~Laloux, P.~Cizeau, J.-P. Bouchaud, and M.~Potters. Noise Dressing of Financial Correlation Matrices. \textit{Physical Review Letters}, 83(7):1467--1470, 1999.
\bibitem[18]{plerou1999} V.~Plerou, P.~Gopikrishnan, B.~Rosenow, L.~A.~N. Amaral, and H.~E. Stanley. Universal and Nonuniversal Properties of Cross Correlations in Financial Time Series. \textit{Physical Review Letters}, 83(7):1471--1474, 1999.
\bibitem[19]{lopezdeprado2018} M.~L\'opez de Prado. \textit{Advances in Financial Machine Learning}. Wiley, 2018.
\bibitem[20]{lopezdeprado} M.~L\'opez de Prado. \textit{Machine Learning for Asset Managers}. Cambridge University Press, 2020.
\bibitem[21]{martin1989} P.~G. Martin and B.~B. McCann. \textit{The Investor's Guide to Fidelity Funds}. Wiley, 1989.
\end{thebibliography}
\end{document}